\documentclass{sig-alternate}

\usepackage{graphicx}
\usepackage{amssymb}
\usepackage{algorithm}
\usepackage{algorithmic}

\usepackage{enumitem}

\newcommand{\mynotex}[1]{}

\begin{document}
\conferenceinfo{MobiArch'12,} {August 22, 2012, Istanbul, Turkey.}
\CopyrightYear{2012}
\crdata{978-1-4503-1526-5/12/08}
\clubpenalty=10000
\widowpenalty = 10000

\title{Enhancing Mobile Data Offloading with Mobility Prediction and Prefetching
}
%
%
%
%
%

\numberofauthors{1} 
%

\author{
\alignauthor
Vasilios A. Siris and Dimitrios Kalyvas \\
\affaddr{Mobile Multimedia Laboratory, Department of Informatics} \\
\affaddr{Athens University of Economics and Business, Greece} \\
\email{vsiris@aueb.gr, kalyvas.d@gmail.com}
}

\maketitle
\begin{abstract}
We present procedures that exploit mobility prediction and prefetching to enhance  offloading of traffic from mobile networks to WiFi hotspots,
for both delay tolerant and delay sensitive traffic.
 We evaluate the  procedures in terms of the percentage of offloaded traffic, the data transfer delay, and the cache size used for prefetching. The evaluation considers empirical measurements and
 shows how various parameters influence the performance of the procedures, and their robustness to time and throughput estimation errors.
\end{abstract}

\category{C.2.1}{Computer-Communication Networks}{Network Architecture and Design}[distributed networks, network communications]

\terms{Algorithms, Performance}

\keywords{Mobile data offloading, cellular and WiFi integration}

\section{Introduction}

Over the last few years, a major trend in mobile networks is the exponential increase of powerful yet affordable personal mobile devices, such as smartphones and tablets, with multiple heterogeneous wireless interfaces that include 3G/4G/LTE and WiFi. The proliferation of such  devices has resulted in a skyrocketing growth of mobile traffic, which in 2011 grew 2.3-fold, more than doubling   for the fourth year in a row, and is expected to grow 18 times from 2011 until 2016\footnote{Source: Cisco Visual Networking Index: Global Mobile Data Traffic Forecast Update, 2011-2016, Feb. 12, 2012}. On the flipside, despite its increase, the mobile data revenue significantly lags behind the exponential growth of data traffic.

One solution to address the strain from the exponential growth of mobile data traffic is network expansion and deployment of 4G/LTE technology, which however is both costly and time consuming.
A second solution is to expand the mobile access network using femtocells, which  exploit the existing backhaul infrastructure (cable, xDSL).
Two issues with femtocells are the interference between femtocells and an operator's metro cells  and the cost of the customer premises devices.

Another alternative is to move a portion of the mobile data traffic to WiFi networks, exploiting the significantly lower cost of WiFi technology and existing backhaul infrastructure.
WiFi offloading can reduce pressure on the most costly component of the mobile network, the RAN (Radio Access Network) which accounts for 70-80\% of the total cost for a mobile network, with the rest of the costs going to the mobile backhaul/core.
The industry has already verified the significance of WiFi offloading by initiatives of large mobile operators: Since May 2010, AT\&T has been deploying WiFi access points in areas with consistently high mobile data traffic. Other major operators including Verizon  and T-Mobile  are also increasing WiFi usage to offload mobile data traffic.

The goal of this paper is to propose procedures that exploit  mobility prediction combined with WiFi and mobile throughput prediction, along with data prefetching to enhance mobile data offloading  to WiFi hotspots. Mobility prediction can provide information on the route that a vehicle will follow and when the vehicle will reach different locations along its route. Such mobility information can be combined with geo-location information regarding WiFi hotspot access and WiFi/mobile throughput, in order to predict the number of WiFi hotspots that the vehicle will encounter,  the duration of access and estimated throughput for each hotspot, and the estimated mobile throughput in route segments where it will have only mobile access.
In summary, our contributions are the following:
\begin{itemize}[topsep=1pt, partopsep=1pt, itemsep=1pt,parsep=1pt]
\item We develop procedures that  exploit mobility prediction for delay tolerant traffic to minimize the data transfer throughput over the mobile network, hence increase the amount of offloaded data traffic.
\item We develop procedures for both delay tolerant and delay sensitive traffic that exploit data prefetching, i.e., proactive caching of data in WiFi hotspots that the vehicle will encounter along its route, to increase the amount of mobile traffic offloaded to WiFi and to reduce the data transfer delay.
\item We evaluate the procedures considering  empirical measurements and a wide range of parameter values, and show their robustness against time and throughput estimation errors.
\end{itemize}
Prior work has shown the predictability of bandwidth for cellular networks  \cite{Yao++08} and for WiFi \cite{Nic++08,Pan++09}. The work of \cite{Yao++09} investigates how bandwidth prediction can improve scheduling in vehicular multi-homed networks and
\cite{Yao++10} investigates improvements for mobile video rate adaptation.
Bandwidth prediction, together with transparent  roaming and handover, for improving video streaming is investigated in  \cite{Eve++11}.
Bandwidth prediction for client-side pre-buffering to improve video streaming is investigated in \cite{Sin++12}. The works \cite{Yao++10,Eve++11,Sin++12} focus on mobile networks, whereas our work investigates mobile data offloading to WiFi. Moreover, unlike \cite{Sin++12} which considers pre-buffering at the client device, we investigate  prefetching data to local caches in WiFi hotspots.

Exploiting  delay tolerance to increase   mobile data offloading to WiFi  is investigated in \cite{Bal++10}. The work of \cite{Lee++10} showed that delay tolerance of up to 100~seconds  provides minimal offloading gains; however, this applies to human daily mobility, rather than vehicles.
The work in \cite{Hou++11} applies a user utility model to reduce the mobile throughput by offloading traffic to WiFi, focusing on a transport layer protocol design to integrate cellular and WiFi networks, and utilizing throughput prediction over a 1-second  interval.
Our work differs in that we consider both delay tolerant and delay sensitive traffic, and exploit data prefetching and  prediction involving multiple WiFi hotpots along a vehicle's route.

The feasibility of using prediction together with prefetching is investigated in \cite{Des++09}, which  develops  a prefetching protocol (based on HTTP range requests), but does not propose or evaluate specific prefetching algorithms. In this paper we propose algorithms for delay tolerant and delay sensitive traffic, and evaluate their performance and robustness against time and throughput estimation errors.
Prefetching to improve the performance of video delivery  is investigated in \cite{Gol++12}, which proposes a centralized model to prefetch data in cellular femtocell networks. Prefetching algorithms to reduce the peak load of mobile networks by offloading traffic to WiFi hotspots are investigated in \cite{Mal++12}. Our work differs in that we consider prefetching for multiple WiFi hotpspots along a vehicle's route, and investigate client-centric algorithms for prefetching in the case of both delay tolerant and delay sensitive traffic.


The rest of this paper is structured as follows:
Section~\ref{sec:route} discusses mobility and bandwidth prediction.
Section~\ref{sec:procedures} proposes procedures to exploit prediction and prefetching for delay tolerant and delay sensitive traffic. Section~\ref{sec:evaluation} evaluates the procedures considering empirical measurements and investigates their robustness to  time and throughput errors. Finally, Section~\ref{sec:conclusions} concludes the paper identifying future research directions.

\mynotex{ Prior/related work on:
\begin{itemize}
\item Offloading of delay tolerant traffic
\item Wifi throughput prediction
\item Mobile throughput prediction, e.g. for streaming multimedia traffic.
\item Prefetching. In space-domain versus time-domain
\item mobile-WiFi handover not the focus of the paper.
\item also how prefetching can be implemented (http get) is discussed in \cite{Des++09}.
\end{itemize}
}

\mynotex{Additional notes:
\begin{itemize}
\item In this paper we investigate prediction in the time-domain, namely a nodes location in some future time. We can also exploit mobility prediction in the space-domain, i.e. to consider multiple future network attachment point where a node can move to, and prefetch data not in one location but in multiple possible future locations.
\end{itemize}

}

\section{Mobility and bandwidth prediction}
\label{sec:route}

Route selection in vehicles is nowadays commonplace.
The proliferation of smartphones equipped with GPS sensors can enhance route selection by periodically sending timestamped geo-location updates to a server (crowd-sourcing). Such timed geo-location data can produce  real-time  travel information for  route segments, which can be used for route selection based on the actual travel time that considers road traffic conditions.
We have implemented and demonstrated such an application, called ``OptiPath'', on Android smartphones \cite{Kal++12}.
Interestingly, the architecture of such an application is very similar to systems that construct bandwidth or connectivity maps for mobile and WiFi access \cite{Nic++08,Yao++09,Pan++09}.

The goal of this paper is not to develop a new system for mobility and bandwidth prediction, but to propose and evaluate procedures that exploit prediction information that is available by systems such as the ones mentioned above, in order to utilize prefetching  and enhance mobile data offloading to WiFi.

\mynotex{ This section should contain:
\begin{itemize}
\item utilization of crowdsourcing
\item note arch is the same
\item text on what additional information is necessary: location of WiFi hotspots, range, throughput, etc.
\end{itemize}
}


\section{Exploiting mobility prediction and prefetching}
\label{sec:procedures}

In this section we present procedures that exploit mobility prediction and prefetching to enhance data offloading from the mobile network to WiFi hotspots. Mobility prediction provides knowledge of how many WiFi hotspots a node (vehicle) will encounter, when they will be encountered, and for how long the node will be in each hotspot's range. In addition to the aforementioned mobility information, we assume that there is information on the estimated throughput that is available in the WiFi hotspots and the mobile network, at different positions along the vehicles's route; for the former, the information includes both the throughput for transferring data from a remote location, e.g., through an ADSL backhaul link, and the throughput for transferring data from a local cache (this throughput estimate is used only in the case of prefetching).

We present procedures for both delay tolerant and delay sensitive traffic. For delay tolerant traffic, the  procedures try to minimize the mobile  throughput  in order to maximize the amount of data offloaded to WiFi, while ensuring that the data is transferred within a given delay constraint.
For delay sensitive traffic the goal is to minimize the transfer delay.

For prefetching, mobility and throughput information is used to estimate the  data to be cached in the next WiFi hotspot that will be encountered.
Prefetching can be advantageous when the throughput of transferring data from a local cache in the WiFi hotspot is higher than the throughput for transferring data from its original remote server location. This occurs when the backhaul link connecting the hotspot to the Internet has low capacity (e.g., for an ADSL link) or when it is congested; this is likely to become more common as the use of the IEEE 802.11n standard increases.

\mynotex{
\begin{itemize}
\item Common feature of both is prefetching, which involves estimating the amount of cached data and the offset within the data object from which to cache.
\item Different is that in the case of delay tolerant traffic, there is a delay threshold within which the data object needs to be transferred. Hence, we try to minimize the throughput of the mobile network. On the other hand, in the case of delay sensitive traffic we use the maximum throughput of the mobile network.
\end{itemize}
}

\subsection{Delay tolerant traffic}

For delay tolerant traffic our objective is to maximize the amount of data offloaded to WiFi, while ensuring that the whole data object is transferred within a given delay threshold.

The pseudocode for the procedure to exploit mobility prediction and prefetching is shown in Algorithm~\ref{alg:delaytolerant}.
The procedure defines the computations and actions that a mobile node takes when it exits a WiFi hotspot, hence has only mobile access (Line~\ref{line:mobile}), and when it enters a WiFi hotspot (Line~\ref{line:wifi}).
Initially, the procedure estimates the amount of data that can be transferred over  WiFi  (Line~\ref{line:datawifi}), and from this the amount of data that needs to be transferred over the mobile network (Line~\ref{line:datatimemobile}). Additionally, the procedure estimates the total time the node has WiFi access (Line~\ref{line:timewifi}) and, from this value and  the delay threshold, it estimates the duration the node has only mobile access (Line~\ref{line:datatimemobile}). From the amount of data that needs to be transferred over the mobile network and the duration of  mobile-only access, the minimum  throughput for transferring data over the mobile network can be estimated (Line~\ref{line:thrmobile}). To perform prefetching,  whenever the node exits a WiFi hotspot the procedure estimates the amount of data to be prefetched (cached) in the next WiFi hotspot (Line~\ref{line:cache}) and the corresponding offset (Line~\ref{line:offset}); this offset depends on the amount of data that will be transferred over the mobile network until the node reaches the next WiFi hotspot.

\newcommand{\swifi}{\mbox{{\scriptsize WiFi}}}
\newcommand{\smobile}{\mbox{{\scriptsize mobile}}}
\newcommand{\sthres}{\mbox{{\scriptsize thres}}}
\newcommand{\smin}{\mbox{{\scriptsize min}}}
\newcommand{\smax}{\mbox{{\scriptsize max}}}
\newcommand{\scache}{\mbox{{\scriptsize cache}}}
\newcommand{\twifi}{\mbox{{\tiny WiFi}}}
\newcommand{\tmobile}{\mbox{{\tiny mobile}}}
\newcommand{\tthres}{\mbox{{\tiny thres}}}
\newcommand{\tmin}{\mbox{{\tiny min}}}
\newcommand{\tmax}{\mbox{{\tiny max}}}
\newcommand{\tcache}{\mbox{{\tiny cache}}}

\newcommand{\tadslno}{\mbox{{\tiny adsl}}}
\newcommand{\tbckhl}{\mbox{{\tiny bckhl}}}

\newcommand{\tnext}{\mbox{\emph {\tiny next}}}
\newcommand{\Offset}{\mbox{\emph {Offset}}}

\renewcommand{\algorithmiccomment}[1]{/* #1 */}

\algsetup{linenosize=\small}

\begin{algorithm}
\caption{Procedure to exploit mobility prediction and prefetching for delay tolerant traffic}
\begin{algorithmic}[1]
\label{alg:delaytolerant}
{\scriptsize
\STATE \textbf{Variables:}
\STATE $D$: size of data object to be transferred
\STATE $T_{\tthres}$: maximum delay threshold for transferring data object
\STATE $N_{\twifi}$: remaining  WiFi hotspots to be encountered until $T_{\tthres}$
\STATE $D^{\tmin}_{\twifi}$: estimated minimum amount of data to be transferred in all WiFi hotspots that will be encountered
\STATE $D_{\tmobile}$: amount of data  to be transferred over mobile network
\STATE $T^{\tmin}_{\twifi, i}, T^{\tmax}_{\twifi, i}$: min, max duration node is connected  to WiFi  $i$
\STATE $T_{\tmobile}$: total duration that node is not in range of WiFi
\STATE $T_{{\mbox{\tiny next WiFi}}}$: average time until node enters range of next WiFi
\STATE $R^{\tmin}_{\twifi, i}, R^{\tmax}_{\twifi, i}$: min, max throughput of WiFi  $i$
\STATE $R_{\tmobile}$: throughput to download data over the mobile network
\STATE $D^{\tcache}_{\twifi, \tnext}$: amount of data cached  in next WiFi hotspot
\STATE $\Offset$: estimated position in data object of data transferred until node enters next WiFi hotspot
\STATE \textbf{Algorithm:}
\IF { node exits WiFi hotspot } \label{line:mobile}
\STATE  $D^{\tmin}_{\twifi}  \leftarrow \sum_{i \in N_{\twifi}} \left ( R^{\tmin}_{\twifi, i} \cdot T^{\tmin}_{\twifi, i} \right )$ \label{line:datawifi}
\STATE $T^{\tmin}_{\twifi}  \leftarrow \sum_{i \in N_{\twifi}} T^{\tmin}_{\twifi, i}$ \label{line:timewifi}
\STATE  $D_{\tmobile} \leftarrow D-D^{\tmin}_{\twifi}$ \& $T_{\tmobile} \leftarrow T_{\tthres}-T^{\tmin}_{\twifi}$ \label{line:datatimemobile}
\STATE $R_{\tmobile} \leftarrow  D_{\tmobile}/T_{\tmobile}$ \label{line:thrmobile}
\STATE $D^{\tcache}_{\twifi, \tnext} \leftarrow  R^{\tmax}_{\twifi, \tnext} \cdot T^{\tmax}_{\twifi, \tnext}$ \label{line:cache}
\STATE $\Offset \leftarrow  R_{\tmobile} \cdot T_{ {\mbox{\tiny  next WiFi}}}$ \label{line:offset}
\STATE Cache $D^{\tcache}_{\twifi, \tnext}$ data in next WiFi starting from $\Offset$
\STATE Transfer data over mobile network with throughput $R_{\tmobile}$
\ELSIF  { node enters WiFi hotspot } \label{line:wifi}
\STATE Transfer data that has not been received up to $\Offset$ from original object location \label{line:adsl_before}
\STATE Transfer data from  local cache
\STATE Use remaining time in WiFi hotspot to transfer data from original object location \label{line:adsl_after}
\ENDIF
\label{line:1end}
\\ }
\end{algorithmic}
\end{algorithm}

Note that when the node enters a WiFi hotspot, it might be missing some portion of the data object up to the offset from which  data has been cached in the hotspot; this can occur if, due to a time estimation error, the node reaches the WiFi hotspot earlier than the time it had initially estimated.
In this case, the missing data needs to be transferred from the data object's original remote location (Line~\ref{line:adsl_before}). Also, again due to a time estimation error, the amount of data cached in the WiFi hotspot can be smaller than the amount the node could have transferred within the time it is in the   hotspot's range. In this  case, the node uses its remaining time in the WiFi hotspot to transfer data from the data object's original location (Line~\ref{line:adsl_after}).

The pseudocode to exploit mobility prediction without prefetching is shown in Algorithm~\ref{alg:delaytolerant_prediction}.
The procedure only estimates the throughput for transferring traffic over the mobile network. When prefetching is not used, the amount of data transferred over WiFi can be smaller, if  the throughput for transferring data from a remote location is smaller than the throughput for transferring data from a local cache. We assume that this is the case, hence Line~\ref{line:dataadsl} of Algorithm~\ref{alg:delaytolerant_prediction} considers the backhaul throughput $R^{\tmin}_{\tbckhl,i}$, rather than the WiFi throughput as in Line~\ref{line:datawifi} of Algorithm~\ref{alg:delaytolerant}.

\begin{algorithm}
\caption{Procedure to exploit mobility prediction  for delay tolerant traffic}
\begin{algorithmic}[1]
\label{alg:delaytolerant_prediction}
{\scriptsize
\STATE \textbf{Variables:}
\STATE $D$: size of data object to be transferred
\STATE $T_{\tthres}$: maximum delay threshold for transferring data object
\STATE $N_{\twifi}$: remaining  WiFi hotspots to be encountered until  $T_{\tthres}$
\STATE $D^{\tmin}_{\twifi}$: estimated minimum amount of data to be transferred in all WiFi hotspots that will be encountered
\STATE $D_{\tmobile}$: amount of data  to be transferred over mobile network
\STATE $T^{\tmin}_{\twifi, i}$: min duration node is connected  to WiFi  $i$
\STATE $T_{\tmobile}$: total duration that node is not in range of WiFi
\STATE $R^{\tmin}_{\tbckhl,i}$: min throughput of backhaul  at hotspot  $i$
\STATE $R_{\tmobile}$: throughput to download data over the mobile network
\STATE \textbf{Algorithm:}
\IF {node exits WiFi hotspot }
\STATE  $D^{\tmin}_{\twifi}  \leftarrow \sum_{i \in N_{\twifi}} \left ( R^{\tmin}_{\tbckhl, i} \cdot T^{\tmin}_{\twifi, i} \right )$ \label{line:dataadsl}
\STATE $T^{\tmin}_{\twifi}  \leftarrow \sum_{i \in N_{\twifi}} T^{\tmin}_{\twifi, i}$
\STATE  $D_{\tmobile} \leftarrow D-D^{\tmin}_{\twifi}$ \& $T_{\tmobile} \leftarrow T_{\tthres}-T^{\tmin}_{\twifi}$
\STATE $R_{\tmobile} \leftarrow  D_{\tmobile}/T_{\tmobile}$
\STATE Transfer data over mobile network with throughput $R_{\tmobile}$
\ELSIF  { node enters WiFi hotspot }
\STATE Transfer data from original object location
\ENDIF
\label{line:1endb}
\\ }
\end{algorithmic}
\end{algorithm}

\mynotex{
\begin{itemize}
\item Key idea is to minimize usage mobile network while assuring that the delay constraint will be satisfied.
\item Must estimate minimum amount of data to be transferred using WiFi. Calculated mobile throughput from remaining data that needs to be transferred.
\item When switch to mobile network do two things: 1) estimate mobile throughput, 2) estimate data to cache in next wifi hotspot.
\item we assume node is connected to mobile network at all times
\item when we do not perform prefetching, the transfer throughput when node is in WiFi hotspot is lower than the throughput of WiFi due e.g. to ADSL link or because throughput for transferring from remote location is smaller
\item In case prefetching is not used, the procedure just decides the throughput to use while connected to the mobile network.
\end{itemize}

}

\subsection{Delay sensitive traffic}

The pseudocode for the procedure to exploit mobility prediction and prefetching in the case of delay sensitive traffic is shown in Algorithm~\ref{alg:delaysensitive}.
Unlike delay tolerant traffic, in order to minimize the transfer delay for delay sensitive traffic, we always use the maximum throughput that is available in the mobile network (Line~\ref{line:offset3}).
As in the case of delay tolerant traffic, when  the mobile node exits a WiFi hotspot it estimates the offset and the amount of data that needs to be prefetched in the next WiFi hotspot that the node will encounter.

\begin{algorithm}
\caption{Procedure to exploit mobility prediction and prefetching for delay sensitive traffic}
\begin{algorithmic}[1]
\label{alg:delaysensitive}
{\scriptsize
\STATE \textbf{Variables:}
\STATE $T^{\tmax}_{\twifi, \tnext}$: maximum duration node is connected  to next WiFi
\STATE $R^{\tmax}_{\twifi, \tnext}$: maximum throughput of next WiFi hotspot
\STATE $T_{{\mbox{\tiny  next WiFi}}}$: time until node enters range of next WiFi
\STATE $R^{\tmax}_{\tmobile}$: maximum throughput of mobile network
\STATE $D^{\tcache}_{\twifi, \tnext}$: amount of data cached  in next WiFi hotspot
\STATE $\Offset$: estimated position in data object of data transferred until node enters next WiFi hotspot
\STATE \textbf{Algorithm:}
\IF {node exits WiFi hotspot }
\STATE $D^{\tcache}_{\twifi, \tnext} \leftarrow  R^{\tmax}_{\twifi, \tnext} \cdot T^{\tmax}_{\twifi, \tnext}$
\STATE $\Offset \leftarrow  R^{\tmax}_{\tmobile} \cdot T_{ {\mbox{\tiny  next WiFi}}}$ \label{line:offset3}
\STATE Cache $D^{\tcache}_{\twifi, \tnext}$ data in next WiFi starting from $\Offset$
\STATE Transfer data over mobile network with throughput $R^{\tmax}_{\tmobile}$
\ELSIF  { node enters WiFi hotspot }
\STATE Transfer data that has not been received up to $\Offset$ from original object location
\STATE Transfer data from  local cache
\STATE Use remaining time in WiFi hotspot to transfer data from original object location
\ENDIF
\label{line:2end}
\\ }
\end{algorithmic}
\end{algorithm}

Note that there is no procedure for exploiting only mobility prediction (without prefetching) for delay sensitive traffic. This is because for delay sensitive traffic the goal is to minimize the transfer delay, hence the maximum available mobile throughput is always used.

\mynotex{
\begin{itemize}
\item Different with procedure in previous subsection is that we use maximum throughput of mobile.
\item Based on mobility prediction need to decide amount of data to cache in next WiFi hotspot and offset of data to cache from
\end{itemize}
}

\section {Evaluation}
\label{sec:evaluation}

\begin{figure}[t]
\centering
\includegraphics[width=2.8in]{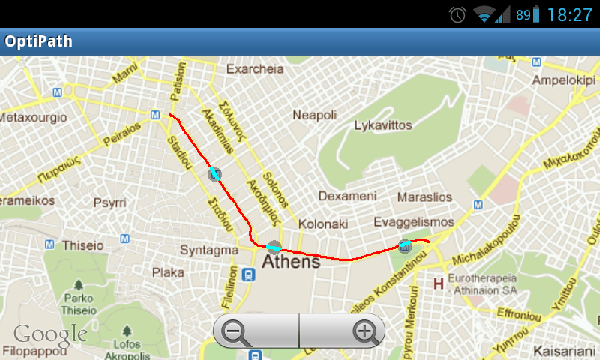}
\caption{\small{Route considered in the evaluation.
The time and location information was obtained empirically. Three WiFi APs where enbedded along the route to conduct the evaluation.
The total travel time for the route is 269~seconds.}}
\label{fig:path}
\end{figure}

The evaluation of the proposed procedures for mobile data offloading considers empirical measurements obtained by the OptiPath application mentioned in Section~\ref{sec:route}. Specifically, we consider a journey between two locations in the center of Athens, Greece, for which the application provides a selected route and mobility information that involves the travel time for different route segments.
Along the route, we embed three WiFi access points, each with range  60~meters. The route considered in our evaluation, with the three embedded  WiFi networks, is shown in Figure~\ref{fig:path} that contains a screenshot of the OptiPath application.
Based on the aforementioned data, we can separate the full route into segments where the moving node has either mobile  or WiFi connectivity, Table~\ref{tab:route_segments}.

\begin{table}[b]
\caption{Connectivity along route. The third column shows the time the node enters the corresponding segment. The times were obtained from empirical measurements and after embedding three WiFi APs  along the route of Figure~\ref{fig:path}.}
    \label{tab:route_segments}
    \centering
 {
        \begin{tabular}{|c|c|c|}
        \hline
        Segment  &  Access &  Time (sec)\\
        \hline \hline
    1 & mobile & 0 \\
    2 & WiFi & 12 \\
    3 & mobile & 32 \\
    4 & WiFi & 149 \\
    5 & mobile & 157 \\
    6 & WiFi & 227 \\
    7 & mobile & 233 \\
        \hline
        \end{tabular}
}
\end{table}

Our goal is to investigate the performance of the proposed algorithms for a wide range of parameter values, including  errors in estimating the duration and the available throughput in each route segment, which can be caused by varying road traffic and wireless link conditions. The parameters  considered  are shown in Table~\ref{tab:values}.
The time error determines how much the times at which the node changes access technology can differ from the empirical values  in  Table~\ref{tab:route_segments}; for example, a 10\% time error means that the time at which the first segment (where the node has mobile access) ends and the  second segment (where it has  WiFi access) begins is in the interval $[0.9 \cdot 12, 1.1 \cdot 12]=[10.8, 13.2]$~seconds. Note that our empirical measurements show that under typical road traffic conditions, the timing for the various route segments can differ 10-20\%.

The throughput error determines the throughput's deviation from its average    in Table~\ref{tab:values}; for example, a 40\% throughput error means that the mobile throughput is in the interval $[0.6,1.4]$~Mbps. Note that our empirical measurements of the mobile throughput for the route in Figure~\ref{fig:path} give values in the range $[1.2,1.4]$~Mbps.
In this paper we only consider the downlink direction, hence the backhaul throughput in Table~\ref{tab:values} refers to the downstream. The improvements achieved in the uplink direction, by uploading data to a local cache,  could be even higher, e.g., in the case of an ADSL backhaul.

The evaluation results presented in the remainder of this section are based on numerically computing the data transferred over the mobile and WiFi networks, for the parameters  in Tables~\ref{tab:route_segments} and \ref{tab:values}.


%
\begin{table}[tb]
\caption{Parameter values.}
    \label{tab:values}
    \centering
 {\small 
        \begin{tabular}{|c|c|}
        \hline
        Parameter  &  Values\\
        \hline \hline
         & 10, 20 (default for delay sensitive), \\
   Data object size      & 30 (default for delay tolerant), \\
         & 40, 50 MB \\
    Mobile throughput  & 1 Mbps (average)\\
    WiFi throughput & 10 Mbps (average)\\
    Backhaul throughput & 2.5, 5 (default), 7.5 Mbps \\
    Time error & 10\% (default), 20\%, 30\%, 40\%  \\
    Throughput error & 20\% (default), 40\%, 60\% \\
        \hline
        \end{tabular}
        }
\end{table}

\begin{figure}[tb]
\begin{center}
\begin{tabular}{c}

\begin{minipage}[b]{0.5\linewidth}
\centering
\hspace{-0.22in}
\includegraphics[width=1.7in] {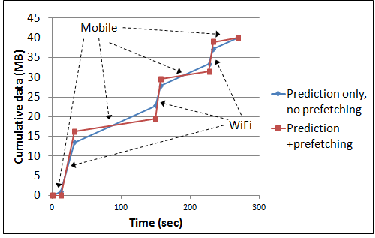}\\
{\footnotesize \small{(a) Delay tolerant}}
\end{minipage}

\begin{minipage}[b]{0.5\linewidth}
\centering
\hspace{-0.22in}
\includegraphics[width=1.7in]{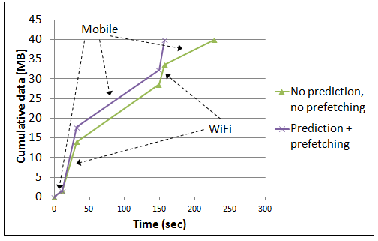}\\
{\footnotesize  \small{(b) Delay sensitive}}
\end{minipage}\

\end{tabular}
\end{center}
\caption[]{\protect \small{Cumulative data as a function of time for delay tolerant and delay sensitive traffic. Data object size: 40~MB.
}}
\label{fig:cumulative}
\end{figure}

The behavior of the procedures of Section~\ref{sec:procedures}  is illustrated in Figure~\ref{fig:cumulative}, which shows the cumulative data transferred as a function of time. Figure~\ref{fig:cumulative} was produced from numerical calculations using the parameters  in Tables~\ref{tab:route_segments} and \ref{tab:values},  assuming that the throughput in each route segment is constant; as a result, the data transferred in each segment is depicted as a straight line.

For delay tolerant traffic, the procedures seek to maximize the amount of traffic offloaded to WiFi, hence to minimize the amount of traffic transferred over the mobile network, while completing the data transfer  until the moving node reaches the end of its journey.
Hence, Figure~\ref{fig:cumulative}(a) shows that for delay tolerant traffic all three WiFi hotspots are utilized. The procedure that exploits prediction and prefetching, Algorithm~\ref{alg:delaytolerant},
differs from the procedure that exploits only prediction, Algorithm~\ref{alg:delaytolerant_prediction}, in that more data is transferred over the WiFi network, which is possible since the transfer throughput from a local cache is higher than the throughput over the hotspot's backhaul link. Similarly, the slopes for the mobile segments are smaller for the procedure that exploits prediction and prefetching, since the amount of data transferred over the mobile network is smaller.

For delay sensitive traffic, Figure~\ref{fig:cumulative}(b), the maximum available mobile throughput  is always used, hence the slope of the mobile segments when prediction and prefetching is used (Algorithm~\ref{alg:delaysensitive}) is the same as when  prediction and  prefetching is not used. As for delay sensitive traffic, prefetching helps offload a larger amount of mobile traffic to WiFi, hence results in a smaller transfer delay compared to the case where no prefetching is used.

In the following subsections we discuss the evaluation results for delay tolerant and delay sensitive traffic. The graphs  show the average values and the 95\% confidence interval from 20 runs of each scenario.
In each run, the time instances that the access technology changes and the throughput of  WiFi, the hotspot backhaul, and the mobile network  is randomly selected based on the average values and error percentages in Tables~\ref{tab:route_segments} and \ref{tab:values}.

\subsection{Delay tolerant traffic}

In this subsection we discuss results for delay tolerant traffic, where a data object needs to be transferred until the  end of the vehicle's route in Figure~\ref{fig:path}. We compare the following three cases: the procedure in Algorithm~\ref{alg:delaytolerant} that exploits mobility prediction and prefetching, the procedure in Algorithm~\ref{alg:delaytolerant_prediction} that exploits only mobility prediction without prefetching, and the case when  prediction is not utilized and the  maximum available mobile throughput  is always used.
The metrics we consider are the percentage of traffic that is offloaded to the mobile network and the cache size when prefetching is used.

\begin{figure}[tb]
\begin{center}
\begin{tabular}{c}

\begin{minipage}[b]{0.5\linewidth}
\centering
\hspace{-0.22in}
\includegraphics[width=1.7in] {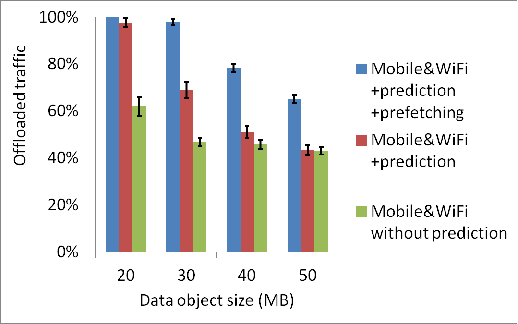}\\
{\footnotesize \small{(a) Data object size}}
\end{minipage}

\begin{minipage}[b]{0.5\linewidth}
\centering
\hspace{-0.22in}
\includegraphics[width=1.7in]{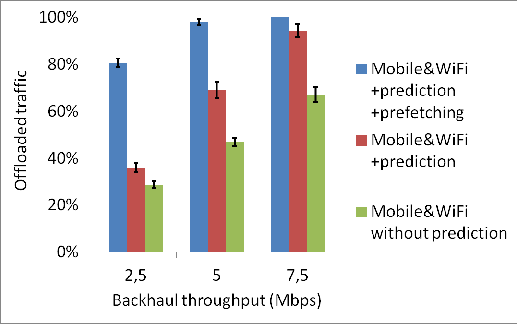}\\
{\footnotesize  \small{(b) Backhaul throughput}}
\end{minipage}\

\end{tabular}
\end{center}
\caption[]{\protect \small{Percentage of offloaded traffic as a function of data object size and backhaul throughput. Delay tolerant traffic.
}}
\label{fig:dt1}
\end{figure}

\medskip
\noindent
\emph{Data object size:} Figure~\ref{fig:dt1}(a) shows the percentage of offloaded traffic for different data object sizes. For all data sizes the performance of the prediction + prefetching scheme is more than 50\% higher compared to the case where  prediction and  prefetching is not used.
For small data object sizes, the performance of both the prediction + prefetching and the prediction schemes is close. The occurs because when the object size is small, then the hotspot backhaul throughput is enough to offload a  high percentage of the mobile traffic, hence the improvement achieved by prefetching and downloading data from a local cache is not significant. On the other hand, for large data sizes, the performance of the prediction scheme is close to the performance when prediction is not used; this occurs because for large object sizes the mobile network is used close to its maximum throughput, hence prediction is not beneficial.

\medskip
\noindent
\emph{Backhaul throughput:} Figure~\ref{fig:dt1}(b) shows the percentage of offloaded traffic for different hotspot backhaul throughputs.
Observe that prefetching can offload   approximately 50\%  (for 7.5 Mbps) and  180\% (for 2.5 Mbps) more mobile traffic compared to when prediction and  prefetching is not used, i.e., when the maximum available mobile throughput is used.
Moreover, when the backhaul throughput is low, the performance when only  prediction is used is close to the performance when  prediction is not used; this happens because when the backhaul throughput is low, then the mobile network needs to be used more, hence the mobile throughput  is close to its maximum.  On the other hand, when the backhaul throughput is high, then the performance of prediction and prefetching is close to the performance when only prediction is used; this occurs because when the backhaul throughput is high and close to the WiFi throughput, there are smaller gains from downloading data from a local cache.

\begin{figure}[tb]
\begin{center}
\begin{tabular}{c}

\begin{minipage}[b]{0.5\linewidth}
\centering
\hspace{-0.22in}
\includegraphics[width=1.7in] {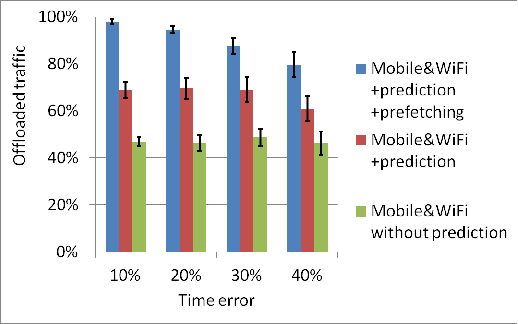}\\
{\footnotesize \small{(a) Time error}}
\end{minipage}

\begin{minipage}[b]{0.5\linewidth}
\centering
\hspace{-0.22in}
\includegraphics[width=1.7in]{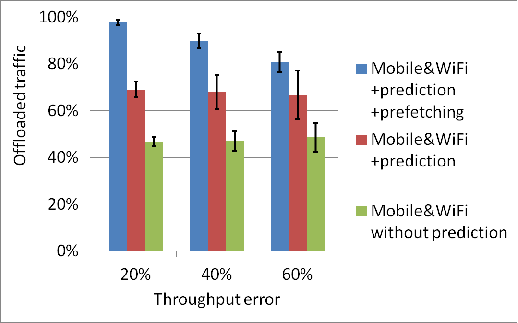}\\
{\footnotesize  \small{(b) Throughput error}}
\end{minipage}\

\end{tabular}
\end{center}
\caption[]{\protect \small{Percentage of offloaded traffic as a function of time and throughput error. Delay tolerant traffic.
}}
\label{fig:dt2}
\end{figure}

\medskip
\noindent
\emph{Time error:} Figure~\ref{fig:dt2}(a) shows how the percentage of offloaded traffic is affected by the time error. Observe that the performance when prediction and prefetching are used decreases as the  time error increases; this occurs because the time error reduces the effectiveness of prefetching. Also observe that the performance when only prediction is used is affected only for 40\% time error.
Nevertheless, the performance when prediction and prefetching are used is more than 70\% higher than when  prediction and prefetching are not used, and more than 30\% higher than when only prediction is used.

\medskip
\noindent
\emph{Throughput error:} Figure~\ref{fig:dt2}(b) shows that the throughput error affects the performance of the prediction and prefetching scheme most. Nevertheless, its performance remains more than 20\% higher than  the prediction-only scheme and more than 65\% higher than the case where prediction and prefetching are not used.

\begin{figure}
\centering
\includegraphics[width=2.0in]{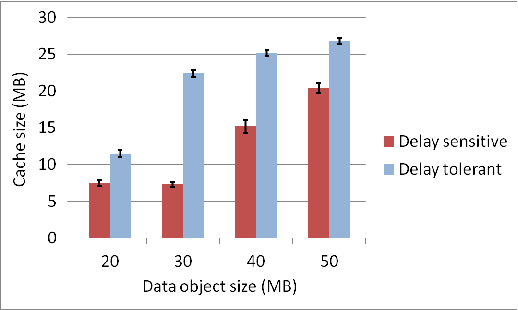}
\caption{\small{Cache requirements for prefetching}}
\label{fig:buffer}
\end{figure}

\medskip
\noindent
\emph{Cache requirements for prefetching:} Figure~\ref{fig:buffer} shows that, as expected, the cache requirements for delay tolerant traffic are higher than for delay sensitive traffic, since  the former's goal  is to maximize the amount of mobile data offloaded to  WiFi. Moreover,  the cache requirements increase with the data  size for delay tolerant traffic. On the other hand, for delay sensitive traffic the cache requirements for  20 and 30~MB data objects remains the same, since for both these sizes only the first WiFi hotspot is used, and  prefetching  is utilized to transfer the same amount of traffic in both cases.

\subsection{Delay sensitive traffic}

In this subsection we discuss the  results for  delay sensitive traffic. A key difference compared to delay tolerant traffic is that now the maximum throughput available in the mobile network is always used.
We compare three cases: the procedure that exploits both mobility prediction and prefetching,  Algorithm~\ref{alg:delaysensitive},  the case where prediction and prefetching are not used, and the case where only the mobile network is used.
The performance metric is the delay for transferring a data object.

\begin{figure}[tb]
\begin{center}
\begin{tabular}{c}

\begin{minipage}[b]{0.5\linewidth}
\centering
\hspace{-0.22in}
\includegraphics[width=1.7in] {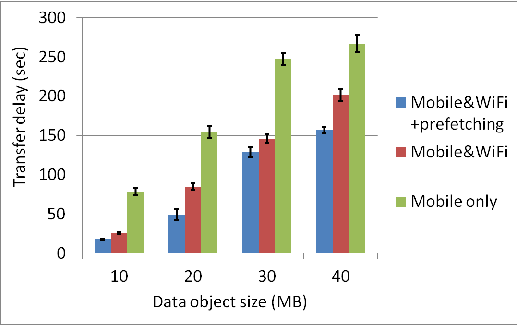}\\
{\footnotesize \small{(a) Data object size}}
\end{minipage}
\begin{minipage}[b]{0.5\linewidth}
\centering
\hspace{-0.22in}
\includegraphics[width=1.7in]{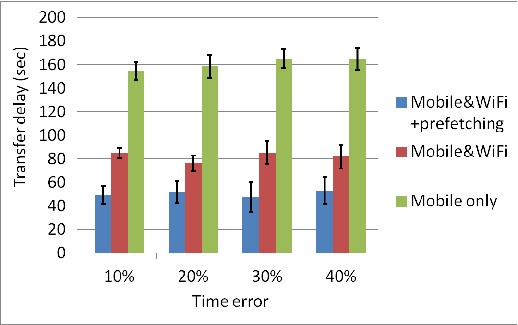}\\
{\footnotesize  \small{(b) Time error}}
\end{minipage}\

\end{tabular}
\end{center}
\caption[]{\protect \small{Transfer delay for different data object sizes and time errors. Delay sensitive traffic.
}}
\label{fig:ds1}
\end{figure}

\medskip
\noindent
\emph{Data object size:} Figure~\ref{fig:ds1}(a) shows the transfer delay as a function of data object size. Prediction and prefetching achieve a transfer delay that is lower by more than 40\% compared to the case where only the mobile network is used.
The procedure that exploits prediction and prefetching achieves transfer delay that is 10\% to more than 40\% lower than the transfer delay when  prediction and  prefetching are not used; this improvement depends on the number of WiFi hotspots that are utilized, which in turn depends both on the location of the WiFi hotspots and the data object size. Specifically, only the first WiFi hotspot is utilized  for data objects with size 20 and 30~MB, and a smaller percentage of the traffic is offloaded for a 30~MB data object; due to this, for a 30~MB object size, the performance when prediction and prefetching is used is closer to  the performance when prediction and prefetching are not used.

\medskip
\noindent
\emph{Time error:} Figure~\ref{fig:ds1}(b) shows the influence of the time error on the transfer delay. Observe that as the time error increases, the variability of the transfer delay increases (the 95\% confidence interval is larger), but the average transfer delay for all schemes remains the same. (We only present present performance results for delay sensitive traffic in the case of time errors; results for throughput estimation errors will be presented in an extended version of this paper.)

\section{Conclusions and Future Work}
\label{sec:conclusions}

We have presented and evaluated procedures that exploit mobility prediction and prefetching to enhance mobile data offloading, for both delay tolerant and delay sensitive traffic. Our evaluation shows how the performance depends on various parameter values, and the robustness of the proposed procedures to time and throughput errors.
Future work includes evaluating the  energy gains  and extending the procedures to allow different tradeoffs between the delay, the amount of offloaded traffic, and the energy efficiency. We are also planning to develop a  prototype to demonstrate the gains of the proposed offloading procedures.

\mynotex{
\begin{itemize}
\item evaluate energy gains
\item implement prototype, extending the OptiPath application. In a first phase this will involve only the procedure to exploit mobility prediction, since prefetching needs support (caching) at the WiFi hotspots.
\end{itemize}
}




\bibliographystyle{abbrv}

{
\bibliography{pref} }

\end{document}